# Investigation User reviews FRO to determine the level of customer loyalty model shahrvand chain stores

**Authors:**

- Mohammad Heydari, "Corresponding Author", Master of Business Administration in International Business student orientation, Payame Noor University, International Center of Assaluyeh, Iran, Postcode: 316814444, E - mail: MohammadHeydari1992@yahoo.com

- Matineh Moghaddam, Master of Business Administration in International Business student orientation, Payame Noor University, Unit of Tehran West, Iran.   E - mail: matineh_Moghaddam@yahoo.com

- Khadijeh Gholami, Master of Business Administration in financial management student orientation, Payame Noor University, Unit of Tehran West, Tehran, Iran. E - mail: egholami_mbi@yahoo.com

- Habibollah Danai, PhD in Business Administration and Professor, Payame Noor University, Tehran, Iran. E - mail: h.danaei@live.com

# Investigation User reviews FRO to determine the level of customer loyalty model shahrvand chain stores

Mohammad Heydari[1], Matineh Moghaddam[2], Khadijeh Gholami[3], Habibollah Danai[4]

**Abstract:**

In this study, focusing on organizations in a rapid-response component model (FRO), the relative importance of each one, from the point of view of customers and their impact on the purchase of Shahrvand chain stores determined to directors and managers of the shops, according to customer needs and their priorities in order to satisfy the customers and take steps to strengthen their competitiveness. For this purpose, all shahrvand chain stores in Tehran currently have 10 stores in different parts of Tehran that have been studied are that of the 10 branches; Five branches were selected. The sampling method is used in this study population with a confidence level of 95% and 8% error; 150 are more specifically typically 30 were studied in each branch. In this study, a standard questionnaire of 26 questions which is used FRO validity using Cronbach's alpha values of "0/95" is obtained. The results showed that each of the six factors on customer loyalty model FRO effective Shahrvand chain stores. The effect of each of the six Foundation FRO customer loyalty model shahrvand is different chain stores.

[1] Master of Business Administration in International Business student orientation, Payame Noor University, International Center of Assaluyeh, Iran. Postcode: 3168144444, Telephone number (09195286393, 021_65350056), E - mail: MohammadHeydari1992@yahoo.com "Corresponding Author"
[2] Master of Business Administration in International Business student orientation, Payame Noor University, Unit of Tehran West, Iran. E - mail: matineh_Moghaddam@yahoo.com
[3] Master of Business Administration in financial management student orientation, Payame Noor University, Unit of Tehran West, Tehran, Iran. E - mail: egholami_mbi@yahoo.com
[4] PhD in Business Administration and Professor, Payame Noor University, Tehran, Iran. E - mail: h.danaei@live.com

## 1. Introduction:

Organizing the distribution of the country's problems, the second program of economic development, reform of distribution networks in order to minimize distribution costs, the number of middlemen and providing judicial protection of consumer rights, including the right to safe and healthy goods and services, the right to All right, has been introduced. So, without the necessary and sufficient customer will certainly fail. Thus, according to the client as one of the vital resources needed to improve the quality and development of services tailored to customers' needs, and desires must constantly be on the agenda of the shahrvand stores. In other words, all software and hardware activities, should represent and reflect the demand and customer expectations. (Arteta B.M., et al. 2000)

Research shows that 5% increase in customer loyalty makes about 25 to 85 percent increase in profits. In addition, the cost of attracting a new customer is between 5 to 11 times to keep an old customer. Customer loyalty is a key to business success, because loyal customers are more profitable, higher repeat purchases, a significant share of the market and the development of others, are introduced.

Today, marketers should ensure that customers are satisfied, they are loyal. Marketing means "growth" and "commitment" of customers in today's perspective; the art of marketing is that the customers of an organization to make the organization's supporters. (Craig Douglas Henry, 2000)

Due to the positive effects of loyalty to the organization and the role of Shahrvand chain stores in the distribution system, be sure about the company's management and personnel, pay attention to this issue and the lack of loyalty or allegiance before buyer's feel is low, factor affecting customer loyalty to the organization identification and reinforcement of the negative impact of these factors can be reduced. The directors of the company must in its marketing strategies and policies, for "customer loyalty" attaches special significance, and every effort in order to maintain customer loyalty to operate. Because failure to do so, leading to customer dissatisfaction and ultimately unsatisfied customer leaves and agencies and issue to the customer dissatisfaction with the other, and leads them to the competitors, so your organization will gradually lose customers and market share losses, are facing. And because the cost of attracting a new customer is high, with increasing aversion customer, the increased costs. These factors lead to a decrease in profitability and ultimately survival is threatened. Studies have also shown that 62% of organizations that customer loyalty is not considered as a necessity and priority, have failed.

It is worth noting that due to the high potential of Shahrvand chain stores in terms of profitability, the cost-benefit analysis and relevant statistics, in the company's profitability and good returns are not achieved, taking into account the fact that loyal customers, make the company more efficient and had a positive effect on long-term profitability and to maintain and develop the company's market share are the managers of the organization should be "customer loyalty" to the panel their plans and continue to the monitor.

Overall, this study seeks to answer the question:

Is the FRO to determine the level of Shahrvand chain stores affects customer loyalty?

**Questions and research hypotheses:**

**Research questions:**

1. Do any one of the six organizations rapid response model (FRO) Shahrvand chain stores affects customer loyalty?
2. Is the effect of each of the six models (FRO) on customer loyalty Shahrvand chain stores, different?

**Research hypotheses:**

1. Each of the six models of rapid-response organizations (FRO) on effective Shahrvand chain stores is customer loyalty.
2. The amount of each of the six models (FRO) on customer loyalty Shahrvand chain stores, are different.

2. **Theory and literature:**
**2.1 theoretical background:**
 - Fast response organizations (FRO): The companies that are able to compete in all six competitive (price, quality, service, time, reliability and flexibility) have a fast response organization referred.
2. Price: The amount paid for providing goods and services to consumers. For example, the amount paid by the consumer to provide the following items: food, clothing, cosmetics and home appliance's chain stores shahrvand. (Giachetti, et al. 2003)
3. Quality: as customers of the characteristics of the product, the quality of different aspects such as performance, compliance, reputable brand, is durable and beautiful.
    - Performance: The performance, usability, comfort and transponder.
    - Durability: Lifetime of durable goods of the said goods.
    - Matching: the degree to which the product conforms to comply with the specifications says.
    - Beauty: suitability and attractiveness beauty goods say.
    - Authentic brand: authentic brand goods.
4. Services and obligations provided all facilities to achieve optimum use of the products that cause customer satisfaction and create value for the customer. The Shahrvand chain stores' services include the appropriate personnel and good humor, appearance of staff, appropriate staff accountability and responsibility of staff commitment, cooling, heating and ventilation inside the store, cleaning, shops, transport facilities from the store to the parking lot or station, complaints and customer feedback, is offering special services. (Dennise L.Duffy, 1995)
5. Time: The time interval between attempts to buy goods and receive goods from Shahrvand chain stores, such as the time available to shop, time to receive the goods needed aeration, aeration time to prepare all the required items, the amount of time employees to service customers receive they and the waiting time to receive and pay at the registers.
6. Reliability: commitment of shahrvand (organization) ethical and legal obligations, such as customer service promised by officials of the Shahrvand chain stores also had a low probability that the product purchased is wrong.
7. Flexibility: the ability of shahrvand to respond to customer needs and adapt to, for example, offers a variety of products in terms of brand and size, kind of different products, providing rapid new products and changes to fit the needs of customers in the chain stores shahrvand. (Noori Clerk, 1999, p. 143)
8. Loyalty: a strong commitment to repurchase a superior product or service referred to in the future, so that the brand or product marketing efforts despite the impact and potential competitors, to be purchased. (Carolyn Folkman curasi, 2002)

FRO competition consisted of:

A rapid-response organization, an organization with a broad vision of the business world and with a range of features and capabilities to deal with turbulence and business environment "advantage" in favor of the company. View new thought for that purpose is required; it should be a new strategic vision beyond traditional systems, support and the new dimension of competition, rather than just considering the "cost of quality" moves, (Sharifi & zhang: 1999). Clearly the factors influencing the decision to purchase, in accordance with a variety of products and a variety of different markets. (Noori, Hamid, 2004; p. 117). However, these factors can be divided into six main groups, namely:

- Product Quality
- Comprehensive support services from suppliers and customers
- Flexibility in product and process
- Strategic use of time as an added value
- The cost of customer orientation and net worth seeing
- Reliability in connection with the fulfillment of the obligations in the market.

The 6 after a rapid-response organization is built.

**2.2 Background Research:**

A study of documents related protocols with the following titles have been carried out:

Evaluation of customer loyalty to the banking system, to investigate the factors influencing consumer loyalty Iranian shampoo and evaluation of experts belonging and loyalty to the organization and its influencing factors. Merely to determine the factors leading to loyalty or loyalty is investigated and loyalty research on the impact of FRO been performed.

## 3. Development of hypotheses and conceptual model:

The above information is provided based on the concept that the design and sequencing of such a model should be considered causal order, and this inference and deduction based on research of the subject developed.

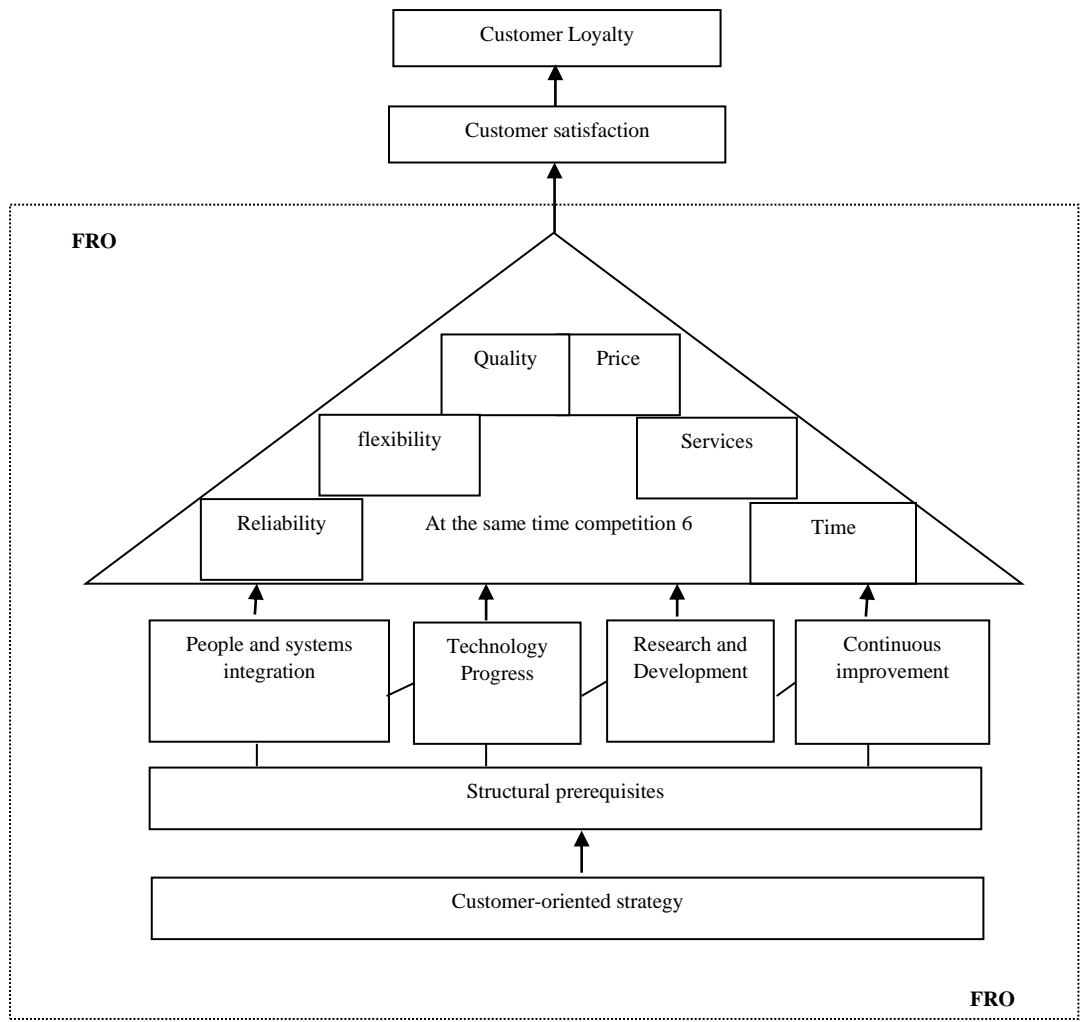

Figure (1): conceptual model

## 4. Methodology:

All the chain stores in Tehran Shahrvand currently have 10 stores in different parts of Tehran that have been studied are that of the 10 branches; Five branches were selected. The sampling method is used in this study population at a confidence level of 95% and 8% error; 150 are more specifically typically 30 were studied in each branch. All purchasers of Shahrvand chain stores in the city of Tehran, the number of samples can be calculated from the formula of the infinite number of sample obtained. Since the population size is unlimited, so the following formula is used to determine the sample size:

$$n = \frac{Z^2_{1-\frac{\alpha}{2}} * p * q}{\Sigma^2} = \frac{(1.96)^2(0.5*0.5)}{(0.08)^2} = 150$$

Table (1): Selected examples in Tehran

| The total number of samples | Selected samples from each branch | Selected branch | Geographical regions of Tehran | Row |
|---|---|---|---|---|
| 150 | 30 | Shahrvand 3 (Behrud) | North | 1 |
|  | 30 | Shahrvand 2 (Baharan) | South | 2 |
|  | 30 | Shahrvand 6 (Jalal aale ahmad) | Center | 3 |
|  | 30 | Shahrvand 1 (Beyhaghi) | East | 4 |
|  | 30 | Shahrvand 8 (Sadeghiyeh) | West | 5 |

In this study, a standard questionnaire of 26 questions which is used FROM validity using Cronbach's alpha values of "0/95" is obtained.

5. **Data analysis:**

**Test research hypotheses:**

(Each of the six models, customer loyalty store chained FRO effective Shahrvand.)

**The first secondary hypothesis**

"Price's intention to repurchase (loyal) customer's effective Shahrvand chain stores."

$$z = = \frac{P_X - P_0}{\sqrt{\frac{P_0 \times q_0}{n}}} \quad \frac{63\% - 50\%}{\sqrt{\frac{50\%(50\%)}{150}}} = 3/18$$

$$d_f = n - 1 \Rightarrow (150 - 1) = 149 \qquad t_{\alpha, df} = t_{0/05, 149} = 1/64$$

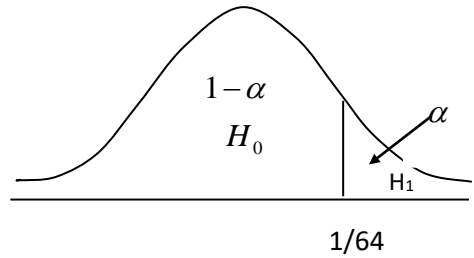

1/64

After comparing test (3/18) with the critical value (1/64) clearly in the H1 test. Thus, it can be said in a confidence level 0/95 H0 does not imply endorsement of the views. Since the expression of a controversial research hypotheses is H0 is then likely to 0/95 percent said the research hypothesis can be confirmed. The "factor prices tend to repurchase (loyalty) customer's effective Shahrvand chain stores."

The second secondary hypothesis

"Quality of intention to repurchase (loyal) customer's effective Shahrvand chain stores."

$$z = \frac{P_X - P_0}{\sqrt{\frac{P_0 \times q_0}{n}}} \qquad \frac{81\% - 50\%}{\sqrt{\frac{50\%(50\%)}{150}}} = 7/59$$

$$d_f = n - 1 \Rightarrow (150 - 1) = 149 \qquad t_{\alpha, df} = t_{0/05, 149} = 1/64$$

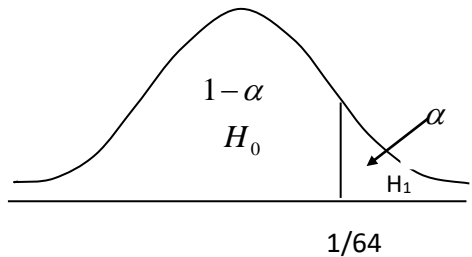

1/64

After comparing test (7/59) with the critical value (1/64) clearly in the H1 test. Thus, a confidence level of 0/95 can be said that observation does not imply endorsement H0. Since H0 is expressing controversial research hypothesis is then likely to 0/95 can be said that research hypotheses were confirmed. The "quality of intention to repurchase (loyal) customer's effective Shahrvand chain stores."

The third secondary hypothesis

"Services in the desire to repurchase (loyal) customer's effective Shahrvand chain stores."

$$z = \frac{P_X - P_0}{\sqrt{\dfrac{P_0 \times q_0}{n}}} \quad \frac{77\% - 50\%}{\sqrt{\dfrac{50\%(50\%)}{150}}} = 6/61$$

$d_f = n - 1 \Rightarrow (150 - 1) = 149$ ، $t_{\alpha, df} = t_{0/05, 149} = 1/64$

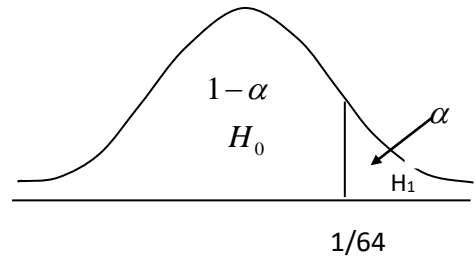

1/64

After comparing test (6/61) with the critical value (1/64) clearly in the H1 test. Thus, a confidence level of 0/95 can be said that observation does not imply endorsement H0. Since the expression of the controversial research hypotheses is H0 is then likely to 0/95 can be said that research hypotheses were confirmed. That 'willingness to purchase services (loyalty) customer's effective Shahrvand chain stores. "

The fourth secondary hypothesis

"Time's intention to repurchase (loyalty) customer's effective Shahrvand chain stores."

According to the data of the first hypothesis z is calculated as follows:

$$z = \frac{P_X - P_0}{\sqrt{\dfrac{P_0 \times q_0}{n}}} \quad \frac{67\% - 50\%}{\sqrt{\dfrac{50\%(50\%)}{150}}} = 4/16$$

$d_f = n - 1 \Rightarrow (150 - 1) = 149$ ، $t_{\alpha, df} = t_{0/05, 149} = 1/64$

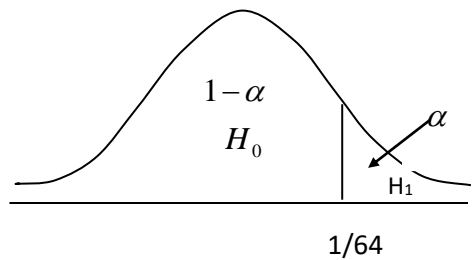

1/64

After comparing test (4/16) with the critical value (1/64) clearly in the H1 test. Thus, a confidence level of 0/95 can be said that observation does not imply endorsement H0. Since the expression of the controversial research hypotheses is H0 is then likely to 0/95 can be said that research hypotheses were confirmed. That is, "the desire to repurchase (loyalty) customer's effective Shahrvand chain stores."

Fifth secondary hypothesis

"Flexibility in the desire to repurchase (loyal) customer's effective Shahrvand chain stores."

$$z = \frac{P_X - P_0}{\sqrt{\frac{P_0 \times q_0}{n}}} \quad \frac{65\% - 50\%}{\sqrt{\frac{50\%(50\%)}{150}}} = 6/12$$

$d_f = n - 1 \Rightarrow (150 - 1) = 149$ ، $t_{\alpha, df} = t_{0/05, 149} = 1/64$

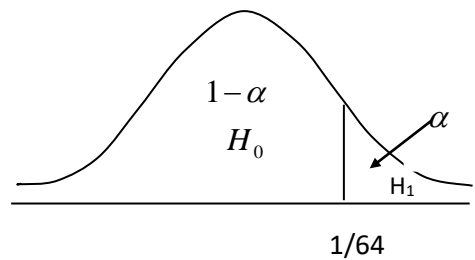

1/64

After comparing test (6/12) with the critical value (1/64) clearly in the H1 test. Thus, a confidence level of 0/95 can be said that observation does not imply endorsement H0. Since the expression of the controversial research hypotheses is H0 is then likely to 0/95 can be said that research hypotheses were confirmed. The "flexibility and willingness to repurchase (loyal) customer's effective Shahrvand chain stores."

Sixth secondary hypothesis

"Reliability repurchase intention (loyalty) customer's effective Shahrvand chain stores."

$$z = \frac{P_X - P_0}{\sqrt{\frac{P_0 \times q_0}{n}}} \quad \frac{84\% - 50\%}{\sqrt{\frac{50\%(50\%)}{150}}} = 8/32$$

$d_f = n - 1 \Rightarrow (150 - 1) = 149$ ، $t_{\alpha, df} = t_{0/05, 149} = 1/64$

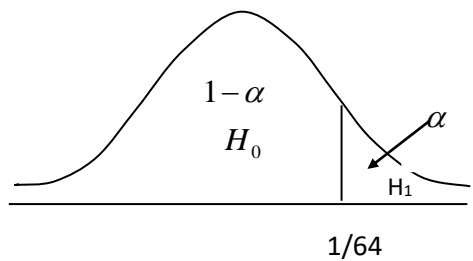

1/64

After comparing test (8/32) with the critical value (1/64) clearly in the H1 test. Thus, a confidence level of 0/95 can be said that observation does not imply endorsement H0. Since the expression of the controversial research hypotheses is H0 is then likely to 0/95 can be said that research hypotheses were confirmed. The "reliability repurchase intention (loyalty) customers' effective Shahrvand chain stores."

**(According to first to sixth sub hypothesis, main hypothesis 1 is confirmed, that each of the six factors on customer loyalty modeled FRO effective Shahrvand chain stores.)**

**The main hypothesis of the second test:**

The effect of six different model Shahrvand FRO customer loyalty chain stores.

The first secondary hypothesis

"The effect of quality on customer loyalty Chain Stores more than price."

$$t = \frac{0.23}{1.077/\sqrt{150}} = \frac{\bar{d}}{S_{\bar{d}}/\sqrt{n}} \quad 2/72$$

$d_f = n - 1 \Rightarrow (106 - 1) = 105$ ، $t_{\alpha, df} = t_{0/05, 105} = 1/64$

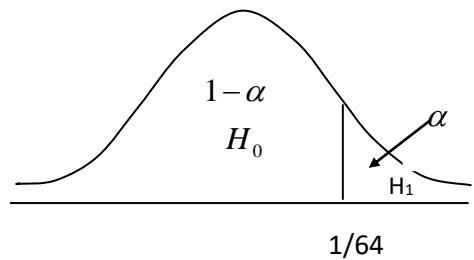

1/64

After comparing test (2/72) with the critical value (1/64) clearly in the H1 test. Thus, a confidence level of 0/95 can be said that observation does not imply endorsement H0. Since H0 is the expression of the controversial research hypotheses is in error after 0/05 can be said that research hypotheses were confirmed. That is, "the impact on customer loyalty Quality Chain Stores more than price."

The second secondary hypothesis

"The effect on customer loyalty chain stores Shahrvand services most of the time."

$$t = \frac{0.14}{0.82/\sqrt{150}} = \frac{\bar{d}}{S_{\bar{d}}/\sqrt{n}} \quad 2/15$$

$$d_f = n - 1 \Rightarrow (106 - 1) = 105 \quad \text{,} \quad t_{\alpha, df} = t_{0/05, 105} = 1/64$$

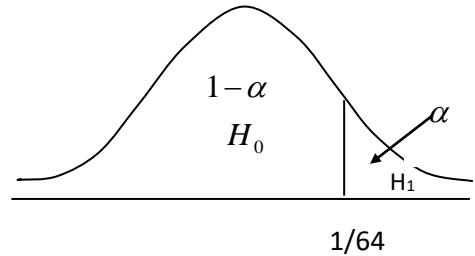

1/64

After comparing test (2/15) with the critical value (1/64) clearly in the H1 test. Thus, a confidence level of 0/95 can be said that observation does not imply endorsement H0. Since H0 is the expression of the controversial research hypotheses is in error after 0/05 can be said that research hypotheses were confirmed. That is, "the impact on customer loyalty services Shahrvand chain stores most of the time."

The third secondary hypothesis

"The effect on customer loyalty chain stores Shahrvand reliability than is flexible."

$$t = \frac{0.48}{1.014/\sqrt{150}} = \frac{\bar{d}}{S_{\bar{d}}/\sqrt{n}} \quad 5/85$$

$$d_f = n - 1 \Rightarrow (106 - 1) = 105 \quad \text{,} \quad t_{\alpha, df} = t_{0/05, 105} = 1/64$$

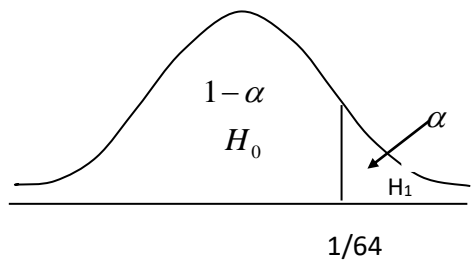

1/64

After comparing test (5/85) with the critical value (1/64) clearly in the H1 test. Thus, a confidence level of 0/95 can be said that observation does not imply endorsement H0. Since H0 is the expression of the controversial research hypotheses is in error after 0/05 can be said that research hypotheses were confirmed. That is, "the impact on customer loyalty chain stores Shahrvand reliability than is flexible."

**(According to the first to third sub-confirmed hypothesis, main hypothesis is confirmed 2; the effects of each of the six Foundation FRO model Shahrvand customer loyalty chain stores are different.)**

### 6. Conclusion:

The results of this study indicate that according to the first to sixth sub hypothesis, main hypothesis 1 is confirmed, that each of the six model Shahrvand FRO effective customer loyalty chain stores. According to the first to third sub-confirmed hypothesis, main hypothesis is confirmed (2), the effect of each of the six Foundation FRO customer loyalty model Shahrvand is different chain stores.

**Offers:**

Increased customer loyalty chain stores all 6 FRO model shahrvand and synergy effects from concurrent attention to every six months, the manager of Shahrvand chain stores is recommended that policies and strategic plans, to consider all aspects and their competitiveness through increased attention to all factors.

Based on the results of the second hypothesis, the team recommended that Shahrvand chain stores, if necessary, and there are limitations in terms of strengthening optimum dimensions, the ratio of quality to price, time and service agent to an agent the reliability of the flexible, pay more attention.

According to the results of the ranking of the six factors to each other, the administrators proposed Shahrvand chain stores, in case of limited resources, hardware and software, for the purpose of allocating resources and customer satisfaction and loyalty in them, consider the following six priorities and the importance of planning and budgeting needed to do. The operating reliability as the most important factor of six in FRO model known Shahrvand chain stores, should be taken into consideration in the allocation of resources to be allocated a larger share.

**References:**

Arteta B.M., R.E. Giachetti, A measure of agility, as the complexity of the enterprise system, Robotics & computer Integrated menufacturing, Volume 20, Issue 6, December 2000 page 495-503.

Carolyn Folkman curasi, Karen Norman Kenedy, from prisoners to Apostels: A Typology of Repeat Buyers & layal customers in servic Bussiness, Journal of services marketing, 16/4, 2002

Clerk Noori production and operations management, translation Dordaneh davari, 1999

Craig Douglas henry, "is customer loyalty a pernicious myth". Newyourk; July 2000.

Dennise L.Duffy, "Customer loyalty strategies" Journal consumer marketing, Vol 15, No 5, 1995, PP 434-448

Giachetti, Ronald E. Luis D.Martinez Oscar A. Saenz, chin-sheng chen, Analysis of the structural measures of flexibility & agility using a measurment theoretical framework International Journal of production Economics, Volume 86, Issue 1, 11 October 2003, pages 47-62

Noori, Hamid, new issues in production and operations management, translation Dordaneh davari, Industrial Management, Vol. I and second, 2004.

Sharifi, H., Z.zhang, A methodology for achieving agility in manufacturing organization: An introduction, International Journal of production Economics, Volume 62, Issue 1-2, 20 may 1999, pages 7-22.